# Crystallisation kinetics of supercooled liquid palladium

**Zuzanna Kostera** * 0009-0002-4729-5523

Warsaw University of Technology Faculty of Physics Koszykowa 75 Warsaw, Poland

**Przemyslaw Dziegielewski** 0000-0002-7544-3960

Warsaw University of Technology Faculty of Physics Warsaw, Poland

przemyslaw.dziegielewski@pw.edu.pl

**Christian Bressler** 0000-0001-9694-5495

European XFEL Holzkoppel 4 22869 Schenefeld, Germany

University of Hamburg Department of Physics Luruper Chaussee 149, 22761 Hamburg, Germany

The Hamburg Centre for Ultrafast Imaging Luruper Chaussee 149, 22761 Hamburg, Germany

**Wojciech Gawelda** 0000-0001-7824-9197

Madrid ES IMDEA Nanociencia Calle Faraday 9, 28049 Madrid, Spain

Adam Mickiewicz University Faculty of Physics Uniwersytetu Poznańskiego 2, 61-614 Poznań, Poland

Paderborn University Faculty of Science, Department of Chemistry Warburger Str. 100 33098 Paderborn, Germany

**Konstantinos Georgarakis** 0000-0003-0918-7310

Cranfield University School of Aerospace, Transport and Manufacturing Cranfield, MK43 0AL, United Kingdom

**Dmitry Khakhulin** 0000-0002-4453-1204

European XFEL Holzkoppel 4 22869 Schenefeld, Germany

**Oleksii I. Liubchenko** 0000-0002-9773-0661

Institute of Physics of the Polish Academy of Sciences al. Lotników 32/46 Warsaw, Poland

**Adam Olczak** 0000-0002-2933-7884

Warsaw University of Technology Faculty of Physics Koszykowa 75 Warsaw, Poland

**Angel Rodríguez-Fernández** 0000-0003-0587-5263

European XFEL Holzkoppel 4 22869 Schenefeld, Germany

**Ryszard Sobierajski** 0000-0003-2580-6900

Institute of Physics of the Polish Academy of Sciences al. Lotników 32/46 Warsaw, Poland

**Klaus Sokolowski-Tinten** 0000-0002-7979-5357

University of Duisburg-Essen, Germany

**Peihao Sun** 0000-0002-3608-1557

Dipartimento di Fisica e Astronomia "Galileo Galilei", Universita degli Studi di Padova Padova 35131, Italy

**Robert W.E. van de Kruijs** 0000-0002-4738-0819

Industrial Focus Group XUV Optics, MESA+Institute for Nanotechnology, University of Twente Drienerlolaan 5 7522 NB Enschede, Netherlands

**Hazem Yousef** 0009-0000-7344-2529

European XFEL Holzkoppel 4 22869 Schenefeld, Germany

**Peter Zalden** 0000-0002-8530-0576

European XFEL Holzkoppel 4 22869 Schenefeld, Germany

**Jerzy Antonowicz** 0000-0002-7781-7540

Warsaw University of Technology Faculty of Physics Koszykowa 75 Warsaw, Poland

Correspondence: * zuzanna.kostera.dokt@pw.edu.pl

## Abstract

In this study, we employ classical molecular dynamics (MD) simulations to investigate the crystallisation kinetics of supercooled liquid palladium and relate the results to time-resolved X-ray diffraction measurements on rapidly quenched Pd thin films. Crystal nucleation and growth rates are determined over the temperature range 700–1150 K (0.38–0.65 $T_m$) by analysing the evolution of the microstructure during the liquid-to-crystal transition. The self-diffusion coefficient of Pd, obtained from the atomic mean-squared displacement, follows Arrhenius behaviour over the investigated temperature range, with an activation energy of 467(6) meV/atom, consistent with available data for supercooled liquid metals. The steady-state homogeneous nucleation rate exhibits a maximum of approximately $4\times10^{35}$ m$^{-3}$ s$^{-1}$ near 0.5 $T_m$. The nucleated grains grow with an effective, population-averaged rate of the order of meters per second. The temperature dependence of this effective rate is better represented by a Wilson–Frenkel expression containing a thermally activated attachment term than by a purely collision-limited expression. Based on multiple statistically independent simulations, a time–temperature–transformation (TTT) diagram for crystallisation onset is constructed. The TTT curve exhibits a nose near 0.5 $T_m$ and 100 ps, corresponding to a critical cooling rate for vitrification on the order of $10^{13}$ K/s. The simulations reproduce the crystallisation onset time and temperature observed in time-resolved X-ray diffraction experiments on optically molten Pd thin films quenched at $5\times10^{11}$ K/s. These results indicate that homogeneous, rather than heterogeneous, nucleation governs the achievable supercooling in the experimentally studied films.

## Keywords

Crystallisation kinetics, Supercooled liquids, Molecular dynamics simulation, Homogeneous nucleation, Crystal growth, Classical nucleation theory (CNT), XFEL, pump–probe, X-ray diffraction

## 1. Introduction

Understanding crystallisation in supercooled liquids (SCLs) is a central problem in materials science and condensed-matter physics. The competition between crystallisation and vitrification governs glass formation and sets important processing limits for many materials. Crystallisation in SCLs is controlled by the interplay between thermodynamic driving forces and the kinetics of atomic motion. As the temperature decreases below the equilibrium melting point, the free-energy difference between the liquid and crystalline phases grows, providing an increasingly strong driving force for crystallisation. At the same time, atomic mobility decreases in the progressively more viscous liquid. The balance between these opposing trends ultimately determines whether the liquid crystallises or instead avoids crystallisation and forms an amorphous structure.

SCL metals generally exhibit very low resistance to crystallisation and crystallise almost instantly when quenched below the equilibrium melting point. Only in alloys with a specifically tuned atomic composition, typically near deep eutectics, a high glass-forming ability can be realized. In contrast, pure single-element metals are an extreme case, with exceptionally high nucleation and growth rates of close-packed crystalline phases in the supercooled state [1,2]. Suppressing crystallisation and thus vitrifying a pure metal is therefore extremely challenging, requiring quenching rates on the order of $10^{14}$ K/s [3]. The degree of supercooling in metallic liquids can be increased through atomization [4], containerless processing [5], or fluxing treatment [6]. The ultimate stability limit of the SCL is set by homogeneous nucleation, which occurs spontaneously in the bulk and, unlike heterogeneous nucleation, cannot be avoided by removing preferential nucleation sites such as container walls or impurities. This distinction has been questioned in pure metals. In MD simulations of SCL iron, the authors [7] observed small satellite grains near larger ones and new grains nucleating on pre-existing grain surfaces, revealing unexpected complexity in the nucleation process and motivating further studies of SCL crystallisation. After the nucleation stage, crystallisation proceeds via rapid propagation of the solid-liquid interface until another grain is reached. As with nucleation, the mechanism of crystal growth in pure metals remains unresolved, and no comprehensive theory of rapid growth exists [8]. The fundamental question of whether the growth rate is diffusion-limited [9,10], governed by thermally activated atomic motion, or collision-limited [11], ultimately constrained by the speed of sound in the liquid, remains open [12].

The ultrashort timescales involved in transitions between the liquid and the solid state make direct experimental investigation extremely challenging, so current understanding largely stems from MD simulations [7,13–18]. Early simulations were constrained by computational resources available at the time, but rapid growth of large-scale MD has since enabled statistically robust measurements of crystallisation kinetics and revealed mesoscale effects, such as non-homogeneous grain distributions [7].

In this work, we utilised MD simulations to study crystal nucleation and growth kinetics in SCL palladium (Pd), an elemental basis for numerous glass-forming metallic alloys. By selecting an optimal system size [19,20] and performing multiple statistically independent runs, we quantified nucleation and growth rates across a wide temperature range, identifying the mechanisms governing crystallisation.

## 2. Methods

**MD simulations**

To accurately represent the crystallisation behaviour of the SCL Pd, an embedded-atom method (EAM) interatomic potential developed specifically for pure Pd was employed in this study [21]. All of the MD simulations were performed using the LAMMPS code [22,23]. Simulations were conducted in the isothermal-isobaric (NPT) ensemble with three-dimensional periodic boundary conditions in a cubic simulation box, a timestep of 2 fs, and no external pressure applied to the system. Both the temperature and pressure were controlled using the Nose-Hoover thermostat and barostat [24]. This setup preserves the correct NPT statistical ensemble and produces well-controlled temperature and pressure trajectories without the velocity rescaling artefacts.

To determine an appropriate simulation system size, a series of preliminary runs was carried out on systems of varying atom counts. The results revealed a pronounced sensitivity of the behaviour of the crystallisation dynamics to system size. Increasing the system size progressively reduced the run-to-run variance in the crystallisation onset time observed in smaller systems. Eventually, a system containing 1,372,000 atoms was selected for the primary simulations, as it yielded consistent, reproducible crystallisation behaviour across independent simulation seeds (different initial coordination structures). Representative results of those preliminary tests are provided in the Supplementary Materials. Moreover, to construct the Time-Temperature-Transformation (TTT) diagram (see the Results section), a complementary series of simulations was performed using a reduced system size of 256,000 atoms. Although individual runs at this size exhibit larger run-to-run variability than those performed on 1,372,000 atoms, the statistical robustness required for the TTT diagram was instead obtained through ensemble averaging. Ten statistically independent simulations were performed at each investigated temperature, and the crystallisation onset time was extracted from their combined distribution. This approach trades single-run reproducibility for ensemble-level statistics, allowing meaningful sampling of nucleation times at each temperature, while keeping the total computational cost tractable.

To investigate the crystallisation process of the SCL, an initial face-centred cubic (FCC) crystalline configuration was first melted and then equilibrated for 100-200 ps at 2500 K, well above the equilibrium melting point of Pd ($T_m$ = 1828 K). This step ensured the complete elimination of any residual crystalline order and the erasure of the system's thermal history. The molten Pd system was then instantaneously quenched by setting the target temperature in the range of $0.38T_m$ to $0.65T_m$ and relaxed at that temperature for approximately 3 ns. Following the step change in temperature, the Nose-Hoover thermostat required on average ~20 ps to stabilize the system temperature to the set value. This thermalisation interval was excluded from the subsequent analysis of nucleation and growth kinetics, for which the origin of time was reset to the end of the stabilisation period.

We visualised and analysed the growth of crystalline structures in SCL Pd, using OVITO [25] software. Nucleation and growth of crystalline grains (here understood as distinct domains of a system in which the atomic lattice has a uniform orientation) were quantified using OVITO's built-in post-processing tools that apply specific structural analysis algorithms to atomistic simulation data. In particular, Polyhedral Template Matching (PTM) [26] was used in conjunction with the Grain Segmentation algorithm. The PTM algorithm distinguishes between ordered and disordered atomic arrangements and classifies order structures according to their crystallographic structure type (e.g. FCC, BCC, HCP). Subsequently, atoms identified as "disordered" by the PTM algorithm, and therefore not belonging to any grain, were removed from the simulation box before applying the Grain Segmentation. The Grain Segmentation, based on the Grain Clustering algorithm, was applied with a minimum grain size arbitrarily set to 100 atoms. This threshold prevented the algorithm from erroneously assigning small disordered regions or noise-like crystalline fragments as separate grains, thereby allowing more reliable observation of individual grain growth. This combined approach enabled detailed, time-resolved observation of grain evolution and the formation of crystal nuclei at the atomic scale. In particular, it allowed crystallites with different structural orientations to be distinguished. This enabled tracking the growth of individual grains, defined as domains with a continuous crystalline lattice. Such grains could be resolved even within evolving grain clusters that other algorithms may identify as single grains.

**XFEL pump-probe experiment**

The time-resolved X-ray powder diffraction (XRD) data analysed in this work were acquired at the European X-ray Free Electron Laser, using femtosecond X-ray pulses following sub-ps optical excitation (pump) at 515 nm wavelength. The measurements were conducted on thin polycrystalline Pd films at the Femtosecond X-ray Experiments (FXE) [27] instrument. The Pd samples consisted of a thin polycrystalline Pd film sandwiched between amorphous cap and substrate layers, the full structural and optical characterisation of which is given in ref. [21]. Because the Pd film thickness was comparable to the optical penetration depth of the pump laser, the absorbed energy was deposited nearly uniformly across the film, producing rapid and homogeneous melting. The cap and substrate then acted as a large thermal reservoir, into which heat was dissipated by phonon thermal conduction, yielding the cooling rates of the order of $5\times10^{11}$ K/s relevant to the present work. Full details of the experimental setup, measurement protocol, thin-film Pd samples and XRD data processing are provided elsewhere [21]. While the experimental results presented in ref. [21] are limited to short pump-probe delay times (up to 30 ps after the optical excitation), covering the initial melting dynamics, the present work includes long-delay-time data corresponding to the quenching stage. By monitoring the transient XRD patterns of the molten film acquired during the quench, in particular the FWHM of the first liquid structure peak, which depends on temperature, and comparing them with XRD patterns calculated for MD snapshots of the liquid (see Supplementary Material for details), we were able to estimate the film temperature and track the time of crystallisation onset.

# 3. Results

Our simulation procedure, described in the Methods section, was conducted in the temperature range of 0.38-0.65$T_m$, yielding a substantial dataset that enables reliable quantification of SCL crystallisation kinetics. Figure 1 provides an overview of the simulated microstructural evolution at three representative temperatures (750 K, 950 K, and 1150 K) and four selected moments of simulations (100 ps, 500 ps, 1500 ps, and 2500 ps) where the system was kept at set temperature. For each temperature-time combination, Figure 1 shows the simulation box visualised in OVITO with the Grain Segmentation algorithm applied. Each visualised system state is accompanied by its corresponding calculated diffraction pattern, with all plots sharing a common intensity (vertical) axis. The XRD patterns are included because, unlike the visualisations of the simulation box, they are representative of the bulk of the system. The presented set of MD snapshots and corresponding XRD patterns provides an overview of the structural evolution from the SCL to the crystalline state, thereby capturing the kinetics of the crystallisation process.

Judging from the volume density of crystalline grains at 100 ps, the nucleation rate is highest at the intermediate temperature of 950 K. Notably, although the onset of nucleation is clearly visible in the MD snapshots, the low crystalline volume fraction prevents its detection in the calculated XRD patterns. A rough analysis of the visualisations reveals that the final (2500 ps) grain structure at 1150 K is significantly coarser than that observed at lower temperatures. This difference is also reflected in the width of the Bragg peaks in the corresponding diffraction patterns, with sharper peaks indicating a larger mean grain size. The differences in the final microstructure are further reflected in the fraction of atoms in a structurally disordered environment, which scales with the grain-boundary contribution. This fraction is approximately 20% for the system crystallising at 1150 K, but increases to about 45% at 750 K (see Supplementary Material).

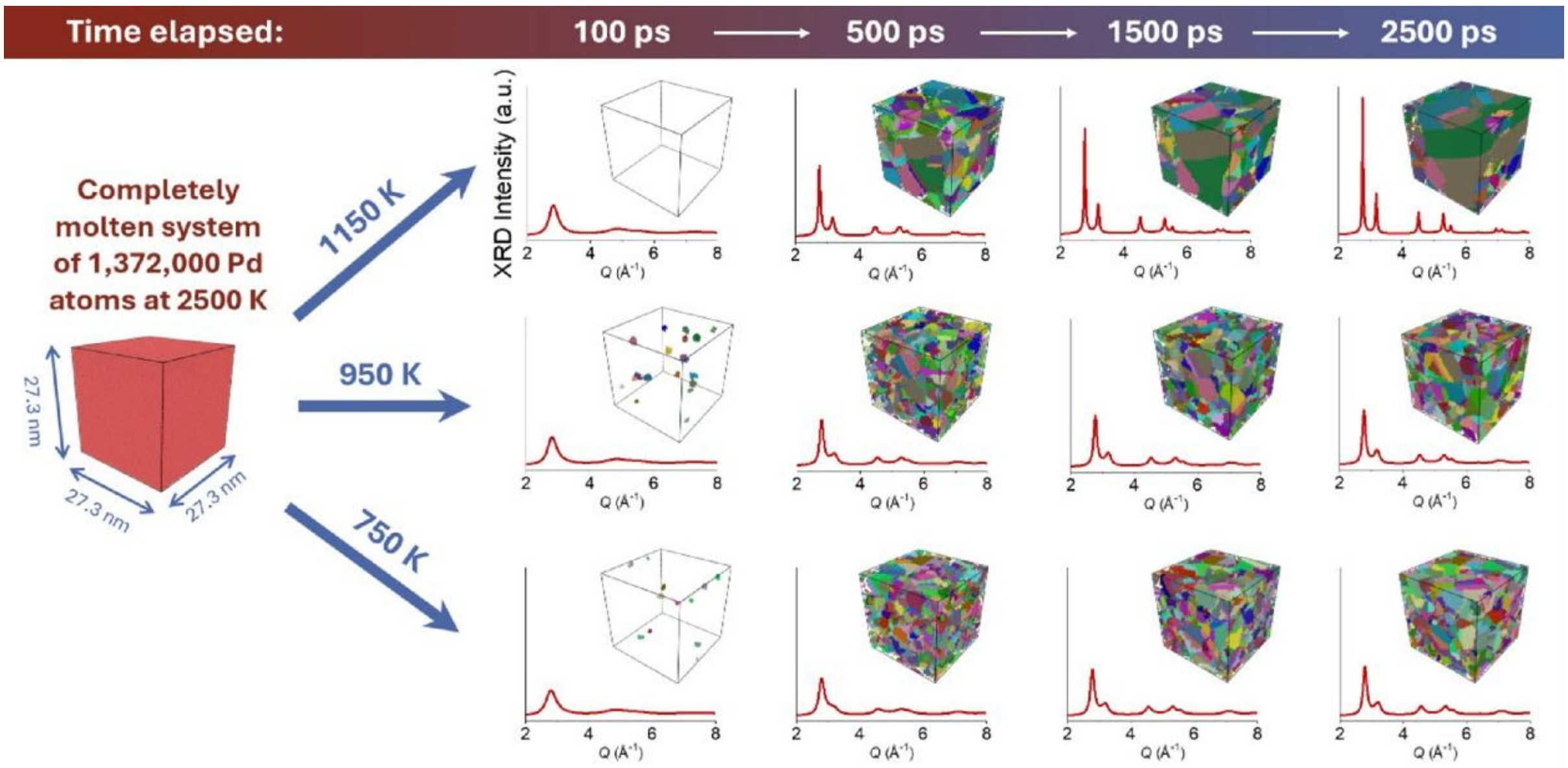

**Figure 1: Selected snapshots of the crystalline microstructure and the corresponding diffraction patterns during the structural evolution of the investigated system of 1,372,000 Pd atoms annealed at 750 K, 950 K, and 1150 K. Grain orientations are colour-coded. For clarity, atoms with a disordered environment were omitted from the visualisation.**

The snapshots shown in Fig. 1 reflect the underlying crystallisation kinetics, namely the temperature dependence of the rates of crystal nucleation $J(T)$ and growth rate $U(T)$. These dependencies arise from the competition between thermodynamic and kinetic factors in the SCL temperature regime. For both crystal nucleation and growth, increasing supercooling enhances the thermodynamic driving force (i.e. the free-energy difference between the liquid and the crystal) but reduces atomic diffusivity. Since sluggish diffusivity is the ultimate factor enabling deep supercooling of a liquid, our first step towards revealing the microscopic mechanism of crystallisation was to quantify atomic mobility from MD structural trajectories. To determine the self-diffusion coefficient $D$ of Pd in the liquid, we tracked the mean-squared atomic displacement (MSD) and used the relation [13,28,29]:

$$D = \frac{1}{2d} \lim_{t\to\infty} \frac{\partial MSD}{\partial t}, \qquad (1)$$

where $t$ denotes time and $d$ is the number of spatial dimensions (3 in the present case). The insert in Figure 2 presents the MSD($t$) dependence for a temperature of 920 K. This temperature was selected as a representative case for the detailed analysis of crystallisation kinetics because it corresponds to the simulated state closest to 0.5 $T_m$ and lies in the vicinity of the nose of the TTT diagram discussed later in this section. After the initial transient ballistic regime of approximately 20 ps, during which atomic motion is inertial, the MSD begins to exhibit a linear dependence on time, indicating the onset of diffusive behaviour. Since the analysis focuses specifically on liquid-state diffusion, the time window of interest was defined as the interval between the end of the ballistic regime and the onset of crystallisation, which occurred at approximately 120 ps. From the slope of the MSD in this time window, the value of $D \approx 1.6 \times 10^{-10}$ m$^2$/s was extracted. As soon as the crystalline phase becomes predominant, the slope of the MSD($t$) curve decreases significantly, since atomic diffusion in the solid is substantially slower than in the liquid. The protocol described above for identifying the appropriate time window for liquid-state diffusion and determining the slope of the MSD($t$) curve was applied at each investigated temperature, yielding the liquid-state diffusion coefficient $D$ as a function of temperature. The results are presented as an Arrhenius plot shown in Figure 2. The linear character of the plot indicates that over the investigated temperature range, $D$ follows an Arrhenius relation

$$D = D_0 exp\left(-\frac{E_a}{k_B T}\right), \qquad (2)$$

where $E_a$ is the activation energy, $D_0$ is the pre-exponential factor, and $k_B$ is the Boltzmann constant. From the linear fit of $ln(D)$ vs $1000/T$ parameters, $E_a$ and $D_0$, were estimated to be approximately 467(6) meV/atom (or 45.1(6) kJ/mol) and 5.8(4) × $10^{-8}$ m$^2$/s, respectively. We note that the slight deviation from linearity in the Arrhenius plot at the lowest temperatures, consistent with a decrease in the activation energy, is accompanied by an increase in the error bars related to the uncertainty in determining the slope of MSD($t$). In our interpretation, this deviation should therefore be considered a numerical artefact.

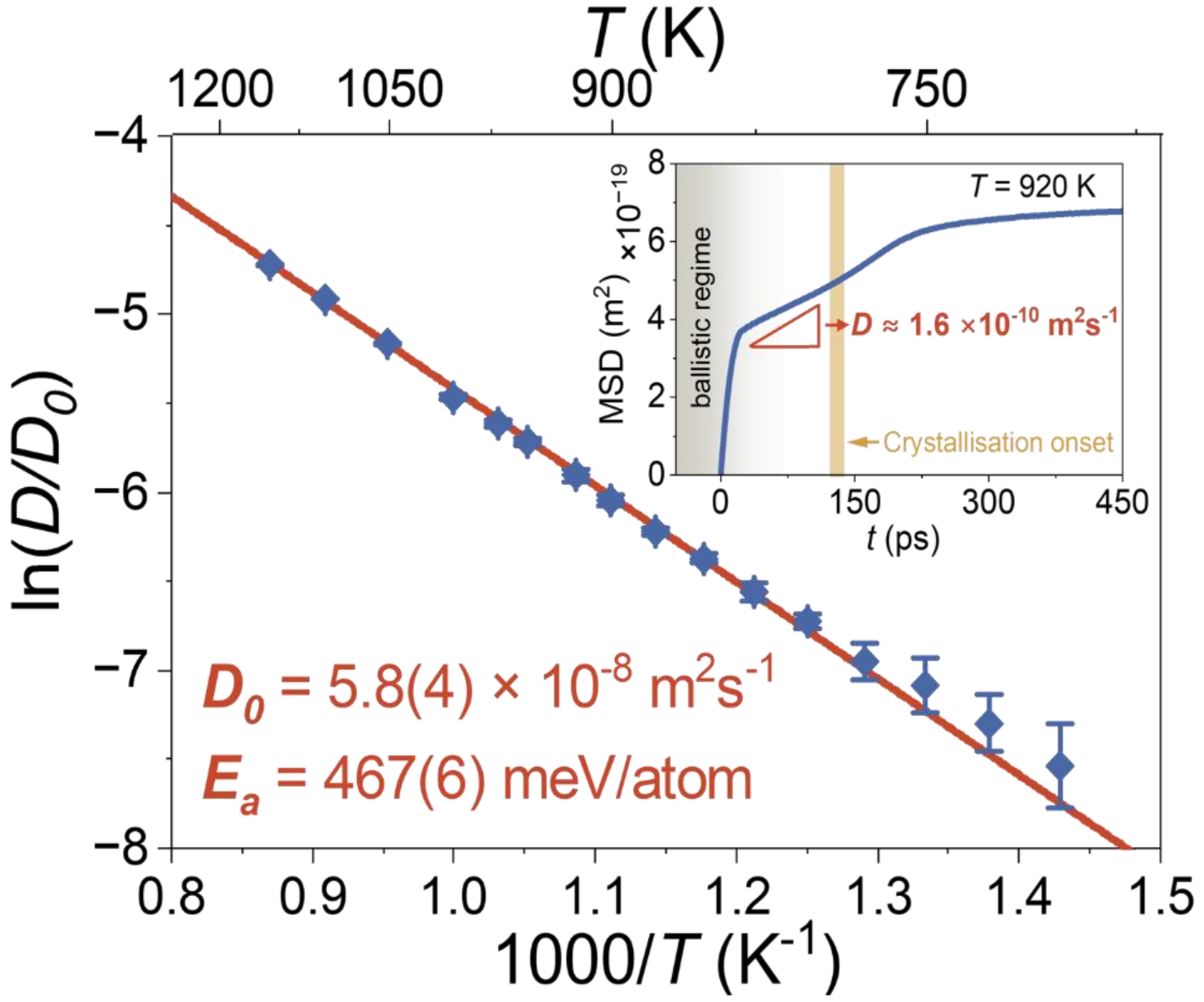

**Figure 2: The Arrhenius plot for the liquid self-diffusion coefficient (symbols) and a linear fit (solid line) with $E_a$ and $D_0$ values. The inset shows the temporal evolution of the MSD at 920 K. The ballistic regime extends over the initial 20 ps, while crystallisation begins at approximately 120 ps. The diffusion coefficient is determined from the slope within this time window.**

In the subsequent stage of the analysis, the crystal nucleation and growth rates were determined. The analysis protocol employed in the present work provides, for each analysed simulation timestep, the total number of crystalline grains present in the simulation box whose size exceeds the prescribed minimum-size threshold, together with the number of atoms comprising each individual grain. It should be noted that the application of such a threshold necessarily precludes direct tracking of the earliest stages of crystalline cluster formation and therefore shifts the apparent nucleation onset towards later times. Nevertheless, this procedure substantially reduces the likelihood of erroneously identifying transient crystalline fluctuations, small ordered clusters, or noise-like structural features as stable grains and thus makes the subsequent analysis more reliable. Furthermore, as demonstrated in the following part of this section, the adopted threshold value of 100 atoms is considerably larger than the estimated critical nucleus size throughout the investigated temperature range. Consequently, the applied protocol restricts the analysis to post-critical nuclei, which are expected to be thermodynamically stable with respect to further growth. In the remainder of this work, such post-critical nuclei are therefore referred to as grains.

In what follows, $U_{eff}$ denotes the effective, population-averaged growth rate obtained from the temporal evolution of the mean equivalent grain radius during the nucleation-and-growth regime, i.e. in the time window between nucleation stage and the onset of grain impingement and subsequent coarsening. We note that $U_{eff}$ should not be interpreted as the local microscopic velocity of the crystal-liquid interface, since continuous nucleation introduces newly formed, smaller grains into the population average, thereby reducing the slope of the mean grain radius relative to the true interface velocity.

Because $U_{eff}$ is defined from the evolution of the mean grain radius, reliable determination of the time interval over which this quantity reflects crystal growth requires accurate identification of the onset of coarsening. Beyond this point, the evolution of the average grain size is no longer controlled by the motion of the liquid-solid interface, but by the reduction of the total grain-boundary surface area, i.e. the minimisation of the total energy of the system. Since polymorphic crystallisation of a monoatomic metal is a purely interface-controlled process [30], limited by the atomic attachment kinetics at the interface rather than by long-range diffusion, the average crystal radius is expected to increase linearly with time, with the slope defining $U_{eff}$. For each grain identified in the simulation, an equivalent spherical radius was computed from the number of atoms it contains. Figure 3b shows the resulting average grain radius as a function of time at $T$= 920 K, corresponding to the data shown in the insert in Fig. 2. The solid line is a linear fit over the interval 110 ps < $t$ < 200 ps, and yields $U_{eff}$ ≈ 2 m/s.

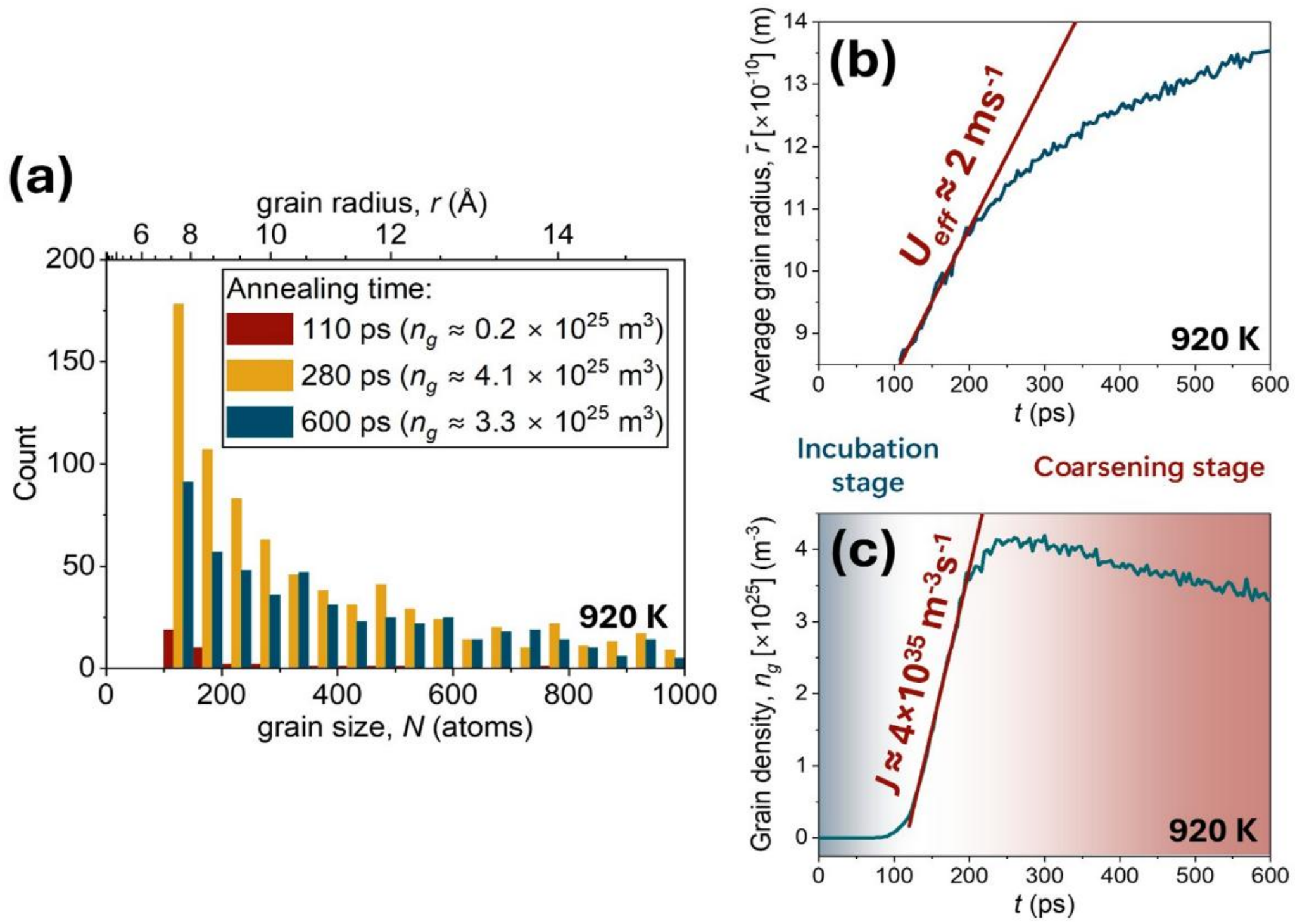

**Figure 3: (a) Histograms of grain sizes $N$ at three different moments (110 ps, 280 ps, 600 ps) of annealing of the system at $T$ = 920 K, with the upper axis of approximated grain radius ($r$) corresponding to the grain size. In the legend attached to each histogram, the grain density ($n_g$) has been indicated. (b) Temporal evolution of the average equivalent grain radius at 920 K. The red solid line represents the linear best fit, and its slope**

**defines the effective population-averaged growth rate $U_{eff}$. (c) Temporal evolution of the grain density at 920 K during the first 600 ps. The incubation and grain coarsening stages are indicated by a coloured background with annotations above the plot. The unshaded region corresponds to steady-state nucleation, where the grain density increases linearly with time. The solid line represents the best-fit line, with the slope indicating the steady-state nucleation rate.**

The nucleation rate can be readily determined from the temporal evolution of the number density of grains identified by the algorithm. Figure 3a presents grain size histograms at three different simulation times. To keep the visibility of the smaller grains present at earlier times, the grain size range shown in Fig. 3a is limited to 1000 atoms. Consequently, many larger grains present at 600 ps fall outside the displayed range. Therefore, it is important to note that the lower histogram bars at 600 ps do not indicate a decrease in the total crystallised volume, but instead reflect the redistribution of crystalline material from smaller to substantially larger grains during coarsening stage. Corresponding grain density ($n_g$) results are shown in Fig. 3c. The dependence exhibits three distinct temporal regimes: an incubation period during which the grain density remains zero, a regime of linear increase corresponding to steady-state nucleation (approximately $4 \times 10^{35}$ $m^{-3}s^{-1}$ for $T$= 920 K), and a slow decline consistent with grain coarsening. Notably, the onset of coarsening inferred from the grain radius data in Fig. 3b (approximately 200 ps) coincides with the upper temporal limit of the linear increase in grain density observed in Fig. 3c, providing additional support for this interpretation.

The procedure described above for 920 K was applied to all investigated temperatures, yielding the temperature dependence of the nucleation and growth rates. Figure 4a presents the derived nucleation rates over the entire investigated temperature range. The nucleation rate exhibits non-monotonic behaviour, reaching a maximum around $4 \times 10^{35}$ $m^{-3}s^{-1}$ at the reduced temperature $T/T_m$ = 0.5. To extract relevant information on the crystal nucleation mechanisms, we interpreted the MD simulation results within the framework of classical nucleation theory (CNT) [2], incorporating the present atomic diffusion data. In applied CNT formulation, atomic diffusion is incorporated through the kinetic factor of the nucleation rate, as the formation and growth of a critical nucleus require atoms to diffuse in the liquid and attach to the crystal interface. The crystal nucleation rate can thus be expressed as

$$J = J_0 \left[ exp\left(-\frac{\Delta G^*}{k_B T}\right) exp\left(-\frac{E_a}{k_B T}\right) \right], \qquad (3)$$

where $J_0$ is the kinetic pre-exponential factor and $\Delta G^*$ is the free energy barrier for nucleation separating the metastable liquid from the stable crystal. By fitting the MD-derived $J(T)$ data points to the above described model and using the previously obtained value of $E_a$ = 45.1(6) kJ/mol, we determined a kinetic factor $J_0$ = 2.3(7)×$10^{41}$ $m^{-3}s^{-1}$. Using the enthalpy of fusion, $\Delta H_f = 17.34(73)$ kJ/mol [31], the liquid–crystal surface energy was estimated as $\gamma \approx 0.17(1)$ J/m$^2$. Assuming a linear dependence of the thermodynamic driving force for crystallisation on supercooling [2], the driving force per atom is $\Delta\mu = \Delta H_f(T_m - T)/(T_m N_A)$ where $N_A$ is Avogadro's number. The corresponding driving force per unit volume is $\Delta g_v = \Delta\mu/\Omega$, where $\Omega$ is the atomic volume of crystalline Pd (approximated as $\approx 1.51 \times 10^{-29}$ m$^3$ per atom). Using these, we calculated the critical nucleus radius $r^*$. The solid line in Fig. 4a represents the resulting $J(T)$ dependence, while Fig. 4b shows the temperature dependence of the critical nucleus radius and the corresponding number of atoms in a critical nucleus. The observed values fall below the arbitrarily selected threshold of 100 atoms, indicating that our algorithm identifies only supercritical grains. More details of the CNT modelling of the simulated nucleation rate are provided in the Supplementary Material.

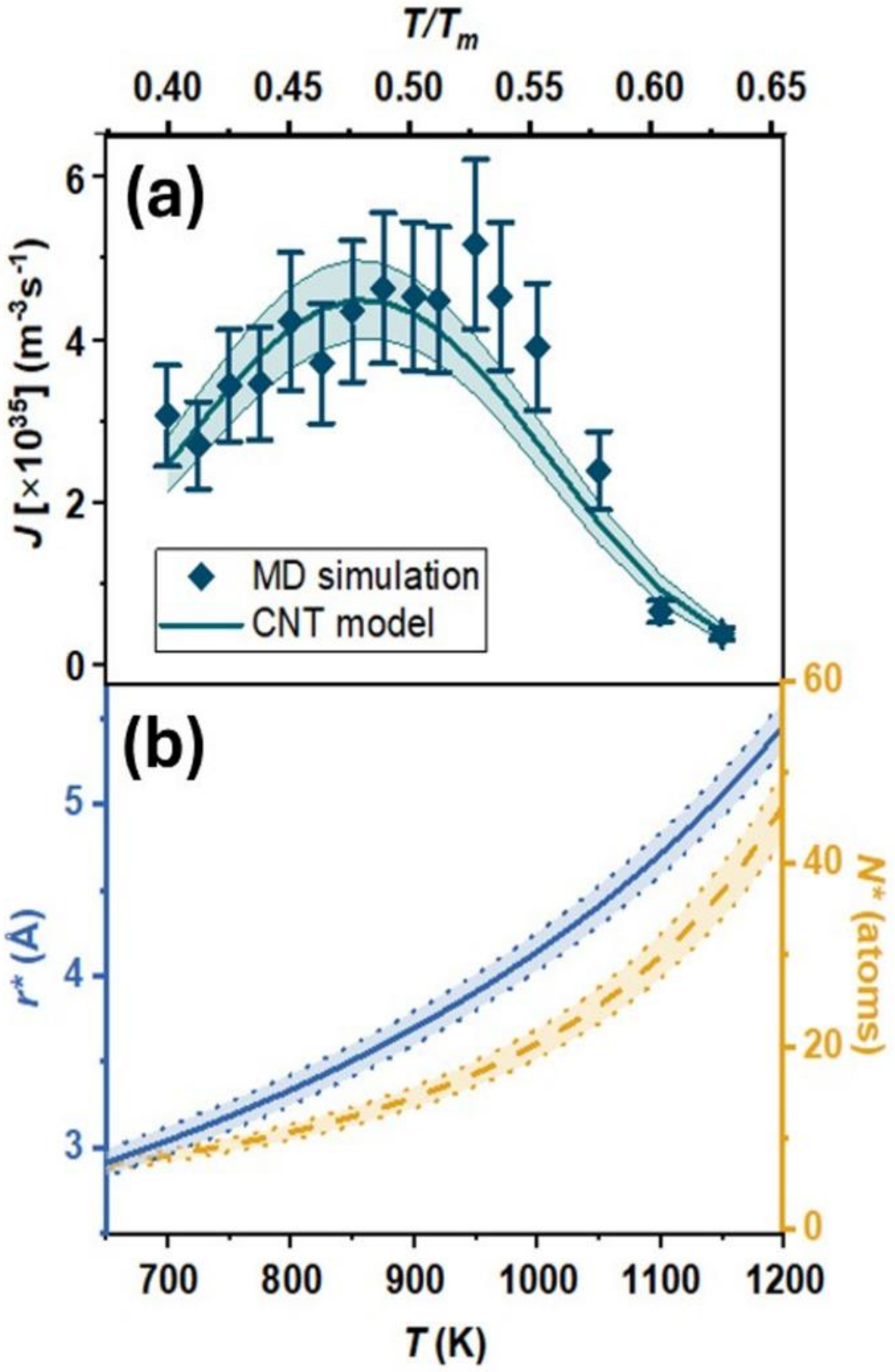

**Figure 4: (a) Temperature-dependent crystal nucleation rates obtained from the MD simulation (symbols) with a fitted CNT model (line). (b) The CNT-derived critical nucleus radius $r^*$ (solid line) compared with the CNT-derived critical nucleus size $N^*$ (dashed line) for the values corresponding to the investigated temperature range.**

Figure 5 plots the crystal growth rate as a function of annealing temperature. Unlike the nucleation rate, the growth rate increases monotonically with temperature, from approximately 1 m/s to 8 m/s over the studied range, with the rate of increase becoming progressively smaller at higher temperatures.

To interpret this behaviour, the MD data were compared with diffusion-limited (Wilson-Frenkel) and collision-limited growth models. First, the simulation results were fitted with the Wilson–Frenkel expression [9,10]:

$$U = f \cdot a \cdot \nu \cdot exp\left(-\frac{E_a}{k_B T}\right) \cdot \left(1 - exp\left(-\frac{\Delta\mu}{k_B T}\right)\right), \qquad (4)$$

where $f$ is the density of growth sites, $a$ is a distance the interface advances when a "liquid atom" manages to attach itself to the crystal, and $\nu$ is a frequency of the order of the Debye frequency. Approximating $a$ by the atomic size of Pd ($a \approx$2.72 Å) and using the expression $\nu_D = \frac{k_B \theta_D}{h}$ (with $h$ standing for Planck's constant and $\theta_D = 274$ K for Pd [32]) to evaluate $\nu$ ( $\sim$ 5.71 THz). The activation energy ($E_a$) used in this expression was previously obtained in aforementioned diffusion analysis by fitting the Arrhenius relation to the self-diffusion coefficients determined from the MSD data. To further investigate the possible grain-growth mechanisms, the predictions of the Wilson-Frenkel model were then compared with the collision-limited growth model, given by [33]:

$$U = f \cdot \frac{a}{\lambda} \cdot \sqrt{\frac{3k_B T}{m}} \cdot \left(1 - exp\left(-\frac{\Delta\mu}{k_B T}\right)\right), \qquad (5)$$

where $\lambda$ is the travel distance of atoms moving at thermal velocity $\sqrt{\frac{3k_B T}{m}}$, and $m$ is the atomic mass. In the general case, $\lambda$ is a fraction of $a$, and thus $\lambda = a$ represents a lower limit for the collision-limited growth rate. For the purpose of qualitatively comparing the temperature dependences predicted by both models with the MD-derived $U_{eff}$ , we assumed the value of parameter $f = 1$. Because the estimated rate is population-averaged, we do not assign this parameter a unique microscopic interpretation. It is retained here as an effective scaling parameter. As shown in the Fig. 5, the realistic parameter set for the Wilson-Frenkel model yields a good fit to both the temperature dependence and the magnitude of the growth rate. We note that the linear approximation of $\Delta\mu$ used in Eq.4 most likely overestimates the true thermodynamic driving force at the largest undercoolings probed here. However, accounting for the non-linearity of $\Delta\mu$ only slightly modifies the $U(T)$ dependence and does not alter the overall character of its temperature variation.

The yellow dashed line in Fig. 5 represents the prediction for the collision-limited growth under the assumption of $\lambda \approx a$ and $f = 1$, deduced from Eq. 5. The resulting values of growth rates corresponding to the collision-limited model span from 250 m/s to 350 m/s across the studied temperature range, significantly exceeding the MD-derived values of $U_{eff}$.

Furthermore, the predictions of the collision-limited to exhibit a temperature dependence qualitatively different from that observed in the MD data. We note, however, that this comparison does not directly identify the elementary attachment mechanism, since the quantity derived from our simulation is a population-averaged growth rate rather than a microscopic interface velocity.

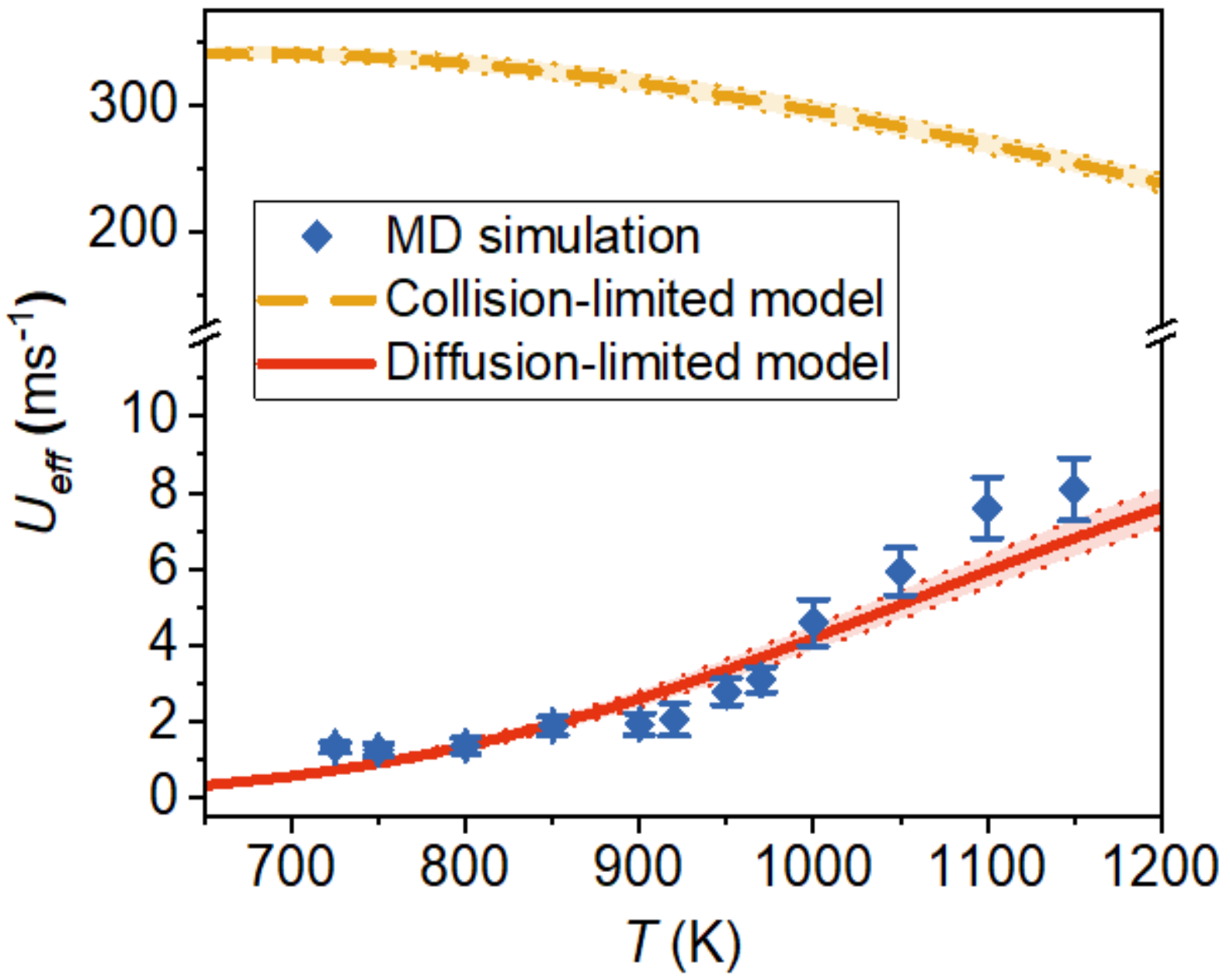

**Figure 5: The effective growth rates $U_{eff}$ determined from the temporal variation of the average equivalent grain radius (solid symbols), shown together with the values predicted by the diffusion-limited (Wilson–Frenkel) model marked by the solid line and the collision-limited growth model (dashed line). The shaded bands represent the uncertainties propagated to the model predictions from the previously determined self-diffusion coefficients (diffusion-limited model only) and from the enthalpy of fusion (both models).**

To validate the predictions of our simulations, the MD results were compared with experimental data on the crystallisation of deeply supercooled thin Pd films. The initially polycrystalline films were melted by a single optical laser pulse and, after a tunable delay time, probed by an X-ray pulse, for which the diffraction pattern in transmission geometry was recorded. Since the interaction of the X-ray pulse with the film resulted in permanent damage (although after the structural information was obtained, so-called "probe-before-destroy" approach), each diffraction pattern was collected from a fresh sample spot. The resulting diffraction patterns, representing transient states of the system, were analysed by deconvolution into contributions from the broad liquid peak and the sharp Bragg peaks of the solid fraction [21]. The crystalline and liquid volume fractions, determined from the relative areas of the corresponding peaks, are plotted in the top panel of Fig. 6a as a function of delay time. The bottom panel shows the corresponding film temperature, estimated from the width of the first peak of the liquid structure factor, following the procedure described in the Methods section. As seen in the figure, the first signatures of the liquid phase appear approximately 3 ps after excitation, at the equilibrium melting temperature (1828 K), and melting is complete by 5 ps, as no crystalline signatures remain. The temperature of the molten film continues to increase until about 10 ps, exceeding 2500 K. Next, cooling sets in as energy transfer from the hot electrons to the lattice is completed and heat dissipates into the cap and substrate layers. The liquid persists for approximately 2.5 ns, during which the temperature decreases to about 1200 K, corresponding to an average cooling rate of $5 \times 10^{11}$ K/s, after which the onset of crystallisation is detected. The dataset presented in Fig. 6 provides an opportunity for a quantitative comparison between the MD-estimated and experimentally measured kinetics of crystallisation of SCL Pd. This comparison can be effectively performed by pinpointing the crystallisation onset on the theoretically predicted TTT diagram shown in Fig. 6b. In this work, the TTT diagram represents the time required for the onset of crystallisation at a given annealing temperature [30]. The resulting C-shaped curve arises from the competition between the thermodynamic driving force and the atomic mobility. The minimum-time point of the curve, known as the "nose", determines the critical cooling rate for glass formation, since only quenches faster than this rate can avoid the crystallisation region and result in vitrification. In Fig. 6b each point in this diagram represents the temperature and time of crystallisation onset obtained from 10 statistically independent simulation runs involving 256,000 atoms, with error bars indicating the standard deviation of the crystallisation onset times. As shown in the figure, the magnitude of the error bars depends strongly on temperature, reflecting the change in the statistical character of crystallisation across the studied range. At lower temperatures, the nucleation rate is high and many supercritical nuclei form nearly simultaneously within each simulation box. Averaging over a number of nucleation events in a single run effectively smooths the inherently stochastic nature of nucleation, so that the crystallisation onset times are reproducible between independent seeds. At higher temperatures, by contrast, the nucleation rate is low and only a few supercritical nuclei appear in each box. Once formed, they grow rapidly because of the higher atomic mobility, so the onset time is essentially set by the appearance of the first critical nucleus, which is a rare and statistically fluctuating event that varies considerably between seeds. Both regimes correspond to nucleation-limited crystallisation. The difference lies in whether the stochastic nature of nucleation is averaged out within each simulation (low T, many events per run) or only across the ensemble of simulations (high T, few events per run). The dense temperature sampling nevertheless allows the TTT curve to be reliably traced across the entire studied range. The position of the nose (approximately 100 ps at $T/T_m \approx 0.5$) yields a critical cooling rate of $10^{13}$ K/s for bypassing crystallisation. This cooling rate exceeds that deduced from the experimental data shown in Fig. 6a by more than an order of magnitude, indicating that the experimental cooling rate is insufficient to avoid crystallisation of Pd, consistent with the observations. Furthermore, the time and temperature of the crystallisation onset predicted by the MD simulations fall within the estimated error bars of the experimental data, providing strong support for the validity of the present modelling approach.

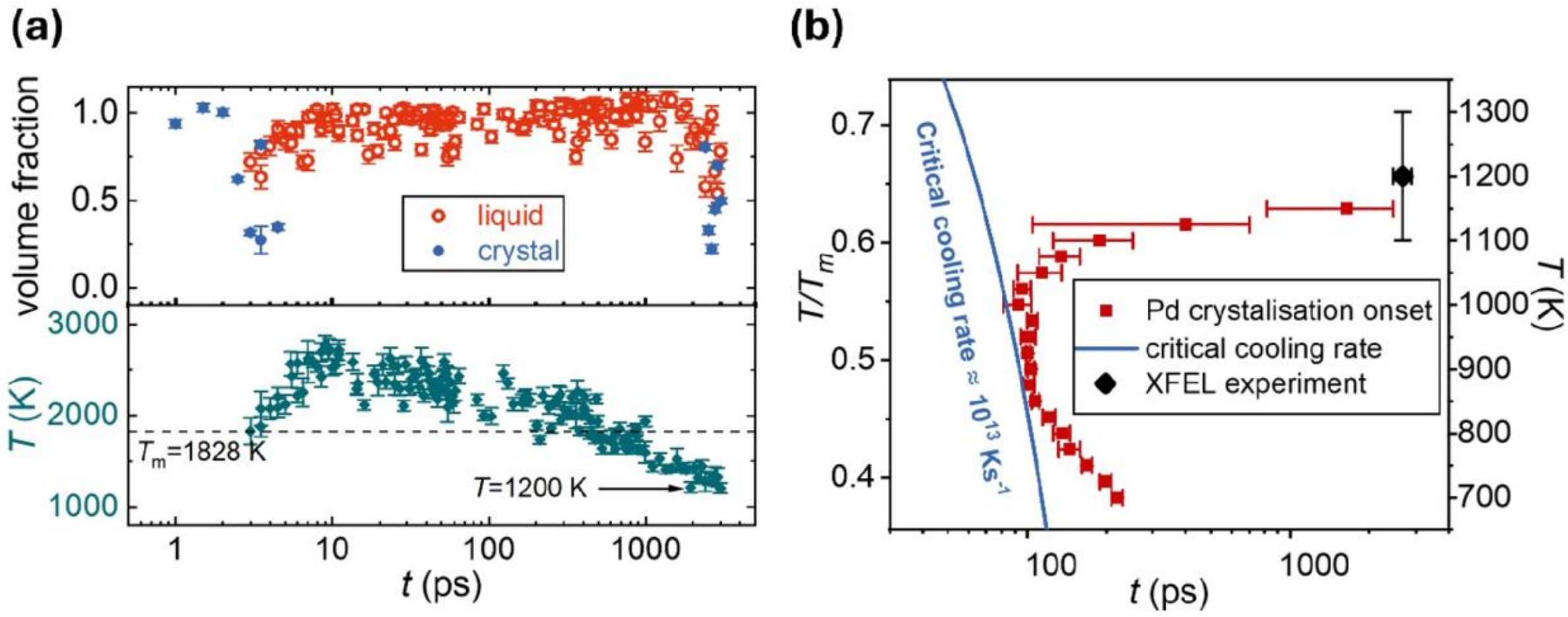

**Figure 6: (a) Upper panel presents volume fraction of liquid (open symbols) and crystalline (full symbols) phase observed in diffraction data of thin Pd films in pump-probe XFEL experiment. Bottom panel presents the temporal relation of temperature during the XFEL experiment, estimated using MD simulations; (b) Time-temperature-transformation (TTT) diagram for the crystallisation onset in SCL Pd. Each data point (full squares) represents the average of 10 statistically independent simulation runs involving 256,000 atoms. Error bars indicate the standard deviation of the crystallisation onset times. The solid line marks the critical cooling rate required to bypass crystallisation. The sole black point on the right side of the graph indicates the crystallisation onset observed in the XFEL experiment (see Fig. 6a).**

# 4. Discussion

In the present work, we developed a realistic numerical model of the crystallisation kinetics of supercooled liquid (SCL) Pd, providing insight into the underlying microscopic mechanisms of the liquid-crystal transition. As noted in the Introduction, Pd serves as an elemental basis for numerous glass-forming metallic alloys, which makes its crystallisation behaviour directly relevant to the broader problem of glass formation in metallic systems. At the same time, Pd is chosen as a model system as it is a representative non-magnetic metal with a moderately high melting point and an FCC equilibrium crystal structure without solid–solid phase transitions. According to the available literature (see [12] and references therein), FCC metals are essentially non-glass-forming systems. Although vitrification of pure BCC metals (e.g. Ta, V) has been demonstrated under extreme cooling rates of approximately $10^{14}$ K/s in nanobridge geometries [3], no comparable result has been reported for pure FCC metals. Significantly, Pd is a primary constituent of numerous metallic glasses, including multicomponent bulk metallic glasses with record-holding critical cooling rates for vitrification below 1 K/s (see, e.g., [34] and [35]). It follows that SCL Pd can be treated as a limiting case of a rapidly crystallising system, i.e. one exhibiting a low free-energy barrier separating the metastable liquid from the equilibrium crystalline state. The height of this barrier, associated with the work of formation of a critical nucleus, can be increased by alloying [36–38] or by reducing the number of heterogeneous nucleation sites by fluxing treatment [6,35].

The ultimate limit for liquid supercooling is homogeneous crystal nucleation, which cannot be eliminated and can only be bypassed by rapid quenching. MD simulations provide a suitable framework for studying homogeneous nucleation, as they inherently exclude heterogeneous nucleation sites and enable direct analysis of atomistic nucleation mechanisms. Early experiments by Turnbull on the solidification of small metal droplets [39], aimed at suppressing heterogeneous nucleation via liquid atomisation, indicate that Pd can be supercooled to approximately $0.8T_m$ (by over 330 K) before homogeneous nucleation sets in. These droplet experiments were performed under near-equilibrium cooling conditions and are therefore not directly comparable to the present MD results, which are limited to nanosecond timescales. The maximum supercooling achieved under near-equilibrium conditions can be interpreted as the "infinite-time" limit of the TTT curve. While extrapolation from nanoseconds to seconds and longer timescales must be treated with caution, we conclude that the present results are consistent (or at least do not contradict) the near-equilibrium droplet experiments in terms of crystallisation kinetics.

A more relevant comparison between the modelling results and the experiment can be made for rapidly quenched thin films, for which the experimental timescale bridges that of the MD simulations. This overlap in timescales provides a unique opportunity for direct validation of the numerical modelling predictions and, in turn, for a physically meaningful interpretation of the experimental results. This comparison yields good agreement between the experimental findings and the theoretical predictions. The crystallisation onset observed at approximately 1200 K corresponds to an undercooling of 620 K ($T/T_m \approx 0.66$). This represents an exceptionally deep undercooling for a liquid FCC metal, close to that achieved in an ultrafast solidification experiment on pure Ag [40]. In terms of interpreting the experimental data, this agreement suggests that rapidly quenched Pd thin films crystallize primarily via homogeneous nucleation and that their supercooling is therefore inevitably limited under the available quenching rates, which are constrained by the thermal properties of the materials involved in the sample (film thickness, heat capacities of the metallic film and substrate/cap layers, and the thermal conductance of their interfaces) [21,41]. In particular, the experimentally accessible cooling rates remain well below the critical cooling rate required to bypass crystallisation, thereby preventing vitrification and favouring crystallisation of the SCL Pd.

The present comprehensive model of crystallisation kinetics provides insight into the underlying transition mechanisms through the magnitude and temperature dependence of key kinetic factors, namely the diffusion coefficient, nucleation rate, and growth rate. According to Fig. 2, the diffusion coefficient exhibits Arrhenius behaviour over the entire investigated temperature range. In particular, no signatures of a crossover indicative of a change in atomic dynamics or a liquid–liquid phase transition are observed [42]. Experimental data on diffusion in pure SCL metals are very scarce, as achieving deep supercooling is extremely challenging, and the available data are largely restricted to temperatures close to the equilibrium melting point. To the best of our knowledge, no experimental data on diffusivity for SCL Pd are available. The existing data on Ni (a group 10 element, like Pd) obtained by quasi-elastic neutron scattering (QENS) [43] indicate an Arrhenius behaviour of $D$ in the range from approximately 200 K above to more than 200 K below the melting point (down to 0.87 $T_m$) with an activation energy of 470 meV/atom and a pre-exponential factor of 7.7 × $10^{-8}$ $m^2/s$. Both the Arrhenius behaviour of $D$ and the magnitude of $E_a$ and $D_0$ are in close agreement with the present results (467(6) meV/atom and 5.8(4) × $10^{-8}$ $m^2/s$, respectively). We also note that the present values are consistent with self-diffusivity data for pure liquid transition metals obtained in the equilibrium liquid temperature range [44].

To further assess the reliability of the present diffusion data, we extrapolated the fitted Arrhenius relation (Eq. 2) to the equilibrium melting point ($T_m = 1828K$), obtaining $D(T_m) \approx 3.0 \times 10^{-9}$ $m^2/s$. This value results from extrapolating approximately 680 K beyond the highest simulated temperature and should therefore be regarded as an estimate rather than a directly simulated quantity. Self-diffusion coefficients of liquid metals at their melting points are known to fall within a narrow range, so that chemically related elements are expected to behave similarly [45]. The reported QENS measurements for different metals at the melting point yield $3.28(7) \times 10^{-9}$ $m^2/s$ for Ni [43], $3.27(4) \times 10^{-9}$ $m^2/s$ for Cu [46] and $2.8(1) \times 10^{-9}$ $m^2/s$ for Fe [47], and our estimate for Pd falls within the same range. Since no experimental data exist for liquid Pd, this agreement cannot serve as a direct validation of the interatomic potential. It does, however, indicate that the atomic mobility predicted here is realistic, which in turn supports the quantitative analysis of nucleation and growth kinetics built upon it.

The nucleation rates of the order of $10^{35}$ $m^{-3}s^{-1}$ reported in the present study are close to those reported for MD-simulated SCL Al [48] . The estimated liquid-crystal surface energy $\gamma$ =0.17(1) $J/m^2$ falls within the range typically observed for metals [2,49,50]. The value reported by Kelton for Pd [2] (0.207 J/m$^2$) is slightly higher but remains in reasonable agreement with the present estimate. In our derivation of the critical nucleus radius, we assume a linear scaling of the thermodynamic driving force for nucleation with temperature. The underlying assumption is that the difference in specific heat capacity between the liquid and the crystal can be neglected. While this approximation is reasonable for pure metals at small supercoolings, it becomes less accurate at the larger supercoolings considered in the present work, and thus $\Delta g_v$ is progressively overestimated at lower temperatures. Since the critical nucleus radius is inversely proportional to $\Delta g_v$, it may therefore be underestimated. The estimated value of $r^*$ for Pd at $0.8T_m$ (see, e.g., [2]) is approximately 1.2 nm, which, considering the limitations discussed above, can be regarded as a reasonable high-temperature extrapolation of the present results.

Whether growth in pure liquid metals is a thermally activated process remains one of the long-standing open questions in solidification theory [10]. The crystallisation front advances into the liquid at velocities far below the speed of sound [51], which represents the limiting case of the collision-limited growth model. The temperature dependence of the effective rates measured here is better represented by a Wilson-Frenkel-type expression containing a thermally activated attachment term than by the purely collision-limited expression, which, evaluated with the same parameters, does not follow their trend. Because the measured quantity is a population-averaged growth rate rather than a microscopic interface velocity, however, this comparison should not be regarded as a direct determination of the elementary attachment mechanism. Nevertheless, the temperature dependence of $U_{eff}$ provides additional information on the factors controlling the effective growth kinetics.

In particular, $U_{eff}$ exhibits a relatively weak temperature dependence at the deepest undercoolings. Between 700 and 900 K, $U_{eff}$ increases only from approximately 1.3 to 2.1 m/s, whereas the bulk self-diffusion coefficient $D$ increases considerably more steeply over the same interval. Some weakening of the temperature dependence is expected within the Wilson–Frenkel picture itself, because upon cooling, the decrease in the kinetic factor is partially compensated by the increasing thermodynamic driving force for crystallisation. However, the substantially weaker variation of $U_{eff}$ compared with $D$ indicates that the kinetic factor governing atomic attachment cannot be identified simply with bulk-liquid self-diffusion.

The effective growth rates obtained here (1-8 m/s) are also markedly lower than the intrinsic interface velocities typically reported for metals from propagation of planar solid-liquid interfaces [52,53]. The comparison, however, involves two different observables. Our values track the time evolution of the mean equivalent grain radius over a population of small crystallites exposing a heterogeneous distribution of local interface orientations, whereas the reference values describe a single, macroscopically flat interface with a specific crystallographic orientation. Two effects associated with this difference are expected to lower the apparent velocity. The first is numerical and related to population averaging: for an approximately constant nucleation rate and a constant individual grain growth rate $U$, averaging over all grains present at a time $t$ gives a mean grain age of $t/2$ and hence yielding $U_{eff} = U/2$ . The second is thermodynamic and related to the Gibbs-Thomson (capillarity) effect, which reduces the bulk (planar interface) driving force $\Delta\mu$ at the curved surface of a grain with radius $r$ by $2\gamma\Omega/r$ [54]. For example, for the grain with radius $2r^*$ the capillary contribution is one-half of the bulk driving force, which, together with population averaging, illustrates how $U_{eff}$ can be several times lower than the planar-interface velocity. Even after accounting for these two effects, the resulting several-fold reduction remains smaller than the difference between the present $U_{eff}$ values and the planar-interface velocities reported for metals, suggesting that additional factors also influence the effective growth rate.

A plausible contribution, kinetic in nature, concerns the local character of the growth front. Freitas and Reed [55] showed that a solid-liquid interface imparts partial ordering on the adjacent liquid, thereby reducing the mobility of nearby atoms and raising the activation energy for attachment above the barrier for bulk liquid diffusion. The effect is strongly anisotropic, with low-index interfaces weakly ordering the liquid and yielding barriers close to the bulk value, whereas rough, high-index interfaces strongly order it and can raise the barrier considerably. The curved surfaces of the small grains studied in our work sample a broad distribution of local orientations and are expected to contain rough, high-index interfaces, whereas a planar-interface calculation probes a single, usually low-index orientation. Since the Wilson-Frenkel expression identifies the attachment barrier with the barrier for bulk liquid diffusion (an identification that interface-induced ordering breaks), this might explain why the simulated growth rates do not scale with the bulk self-diffusion coefficient. A related conclusion was reached by Hu and Tanaka [56] from simulations of a metallic alloy, in which liquid preordering was found to affect not only nucleation but also growth through its effect on the liquid-crystal interfacial energy. On that basis the authors [56] argue that the growth rate is not determined by the diffusion coefficient and the driving force alone, as assumed in the Wilson–Frenkel description. The interface-induced ordering mechanism proposed by Freitas and Ree [55] provides a possible microscopic explanation for the weak temperature dependence $U_{eff}$at low temperatures, as changes in the structure of the interfacial liquid may alter the activation barrier for atomic attachment and weaken its direct coupling to bulk-liquid diffusion. However, we emphasize that interface-induced ordering, local activation barriers and interface roughness have not been determined for Pd nuclei in the present work.

# 5. Conclusions

This work presents a comprehensive MD-based model of crystallisation in supercooled liquid Pd, which shows good agreement with time-resolved X-ray diffraction data for rapidly quenched thin Pd films and employs physically realistic parameters. The model indicates that homogeneous crystal nucleation limits the supercooling of liquid Pd to about 0.66 $T_m$ even under ultrafast quenching conditions of $5 \times 10^{11}$ K/s. The determined effective growth rates indicate that thermally activated atomic attachment is the growth-limiting factor. The results further suggest that homogeneous, rather than heterogeneous, nucleation governs the achievable supercooling in the experimentally studied films. Overall, these findings provide a consistent framework for interpreting and predicting crystallisation behaviour in supercooled metallic systems under extreme quenching conditions.

# Funding sources

This work was supported by the National Science Centre, Poland, grant agreement No 2021/43/B/ST5/02480, and the Deutsche Forschungsgemeinschaft (DFG, German Research Foundation) through Project 278162697-SFB 1242.

# Acknowledgements

This research was carried out with the support of the Interdisciplinary Centre for Mathematical and Computational Modelling, University of Warsaw (ICM UW), under computation allocations No. G98-1896, G99-2123 and G101-2359. We acknowledge European XFEL in Schenefeld, Germany, for provision of X-ray free-electron laser beamtime at the Scientific Instrument FXE (Femtosecond X-Ray Experiments) and would like to thank the staff for their assistance. The access to the European XFEL was financially supported by a grant of the Polish Ministry of Education and Science - decision no. 2022/WK/13.

# Declaration of generative AI use

During the preparation of this work, the author(s) used Claude (Anthropic) and ChatGPT in order to improve the language and readability of selected passages of the manuscript. After using this tool, the author(s) reviewed and edited the content as needed and take full responsibility for the content of the published article.

# Supplementary materials

## 1. Preliminary test – system size

Each data point represents the mean nucleation onset time for a given annealing temperature *T* and system size, averaged over 10 simulation runs with distinct initial configurations. Error bars indicate the standard deviation.

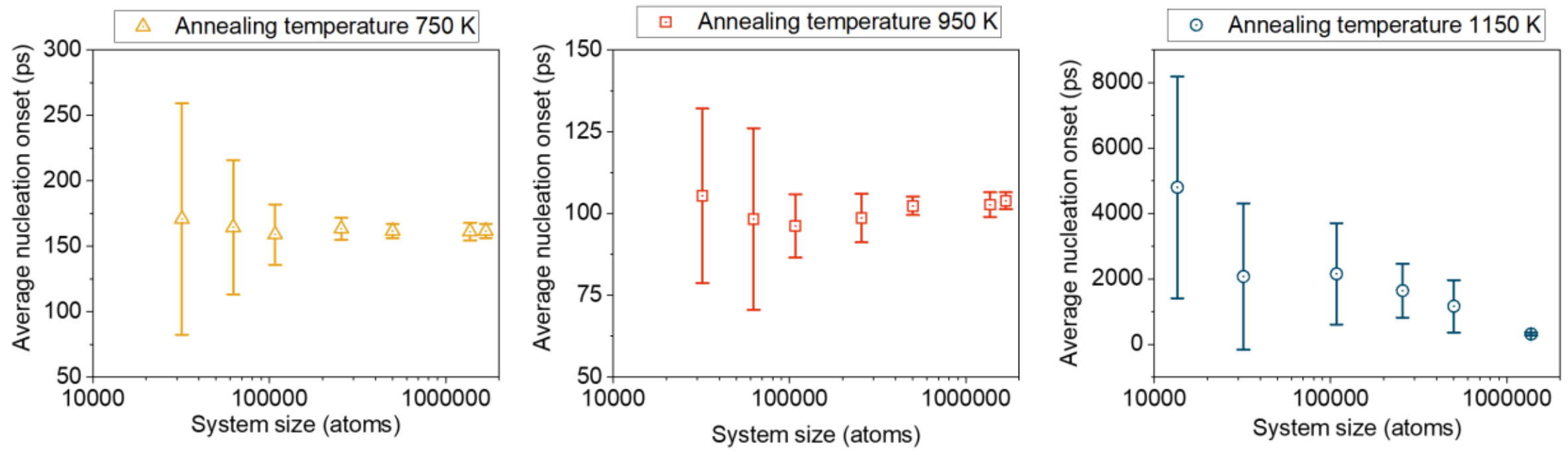

## 2. Classical Nucleation Theory

To analyse the temperature dependence of the homogeneous nucleation rate, we employed the Classical Nucleation Theory (CNT) framework. In this section, we show how the standard CNT expression for the nucleation rate was transformed into the linear fitting form used in the present work. Firstly, we take a steady-state homogeneous nucleation rate written as:

$$J = J_0 \exp\left(-\frac{E_a}{kT}\right) exp\left(-\frac{\Delta G^*}{kT}\right),$$

where J -nucleation rate, $J_0$ – a temperature-independent prefactor, $E_a$– activation energy associated with atomic transport, k – Boltzmann's constant, T – temperature, and $\Delta G^*$ – critical free-energy barrier for nucleation. In the present work, $E_a$ was taken from the Arrhenius analysis shown in the main article and was therefore treated as an independently known parameter.

Within CNT, the free-energy barrier for the formation of a spherical critical nucleus is given by

$$\Delta G^* = \frac{16\pi\gamma^3}{3(\Delta g_v)^2},$$

Where γ is the solid-liquid interfacial free energy and $\Delta g_v$is the volumetric Gibbs free-energy difference between the parent phase and nucleating phase. For solidification, $\Delta g_v{}^*$ can be approximated by the standard linearized form

$$\Delta g_v \approx \frac{\Delta H_f \Delta T}{V_m T_m},$$

where $\Delta H_f$ is molar enthalpy of fusion, $\Delta T = (T_m - T)$ is the undercooling, $T_m$ is the melting temperature and $V_m$ is the molar volume. The approximation above assumes linear scaling of the thermodynamic driving force with the undercooling. Then we can substitute:

$$\Delta G^* = \frac{16\pi\gamma^3 T_m^2 V_m^2}{3\Delta H_f^2 (T_m - T)^2}.$$

Then we defined a new variable $A = \frac{16\pi\gamma^3 T_m^2 V_m^2}{3k\Delta H_f^2}$ , and substitute it into the nucleation rate expression:

$$J = J_0 exp\left(-\frac{E_a}{kT}\right) exp\left(-\frac{A}{T(T_m - T)^2}\right).$$

Then we calculated the logarithm of this expression:

$$lnJ + \frac{E_a}{kT} = lnJ_0 - \frac{A}{T(T_m - T)^2}.$$

In the next step, we can construct a plot where:

$$Y = lnJ + \frac{E_a}{kT} \; and \; X = \frac{1}{T(T_m - T)^2},$$

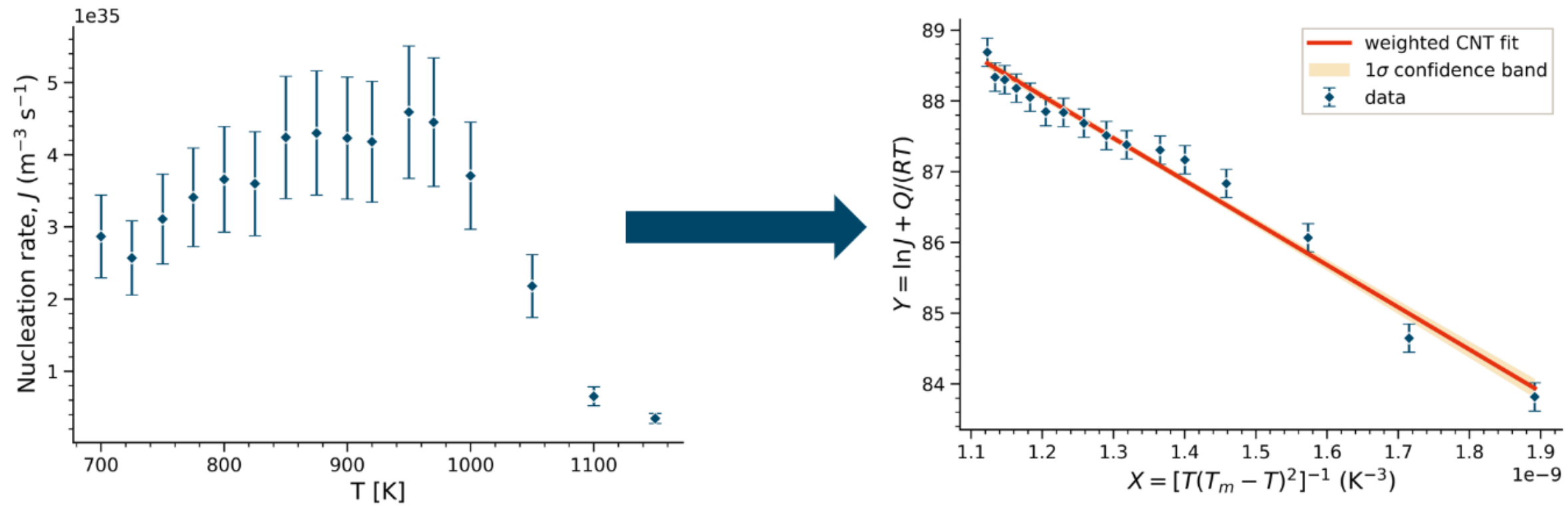

that can be fitted by linear expression $Y = aX + b$, where $a = -A$ and $b = lnJ_0$ .
The interfacial free energy is then calculated from:

$$\gamma = \sqrt[3]{\frac{3kA\Delta H_f^2}{16\pi T_m^2 V_m^2}},$$

where molar volume is given by expression:

$$V_m = \frac{M}{\rho},$$

$M$ – molar mass and $\rho$ – mass density.

The temperature-dependent barrier, driving force, and critical radius are eventually obtained from derived expressions, yielding results:

$$J_0 = 2.31(71) \cdot 10^{41} \frac{m^3}{s}$$

$$\gamma = 0.1812(23)\,\frac{J}{m^2}.$$

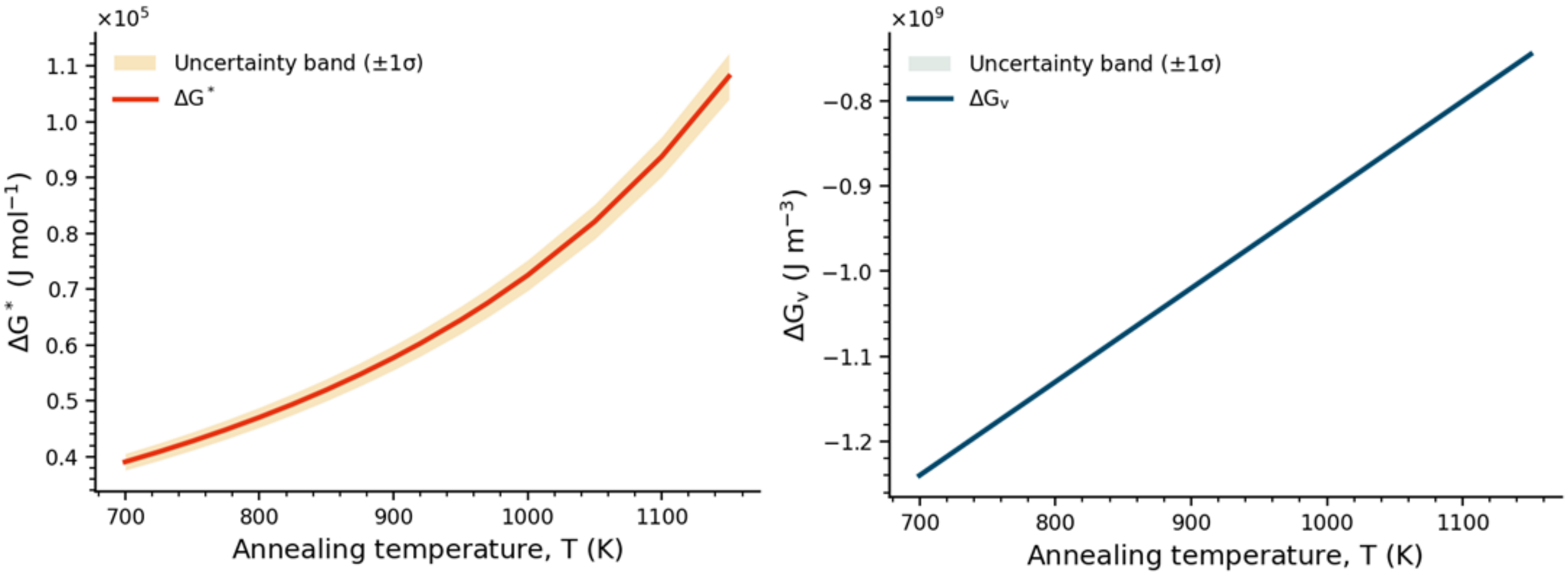

## 3. Temperature estimation for experimental data

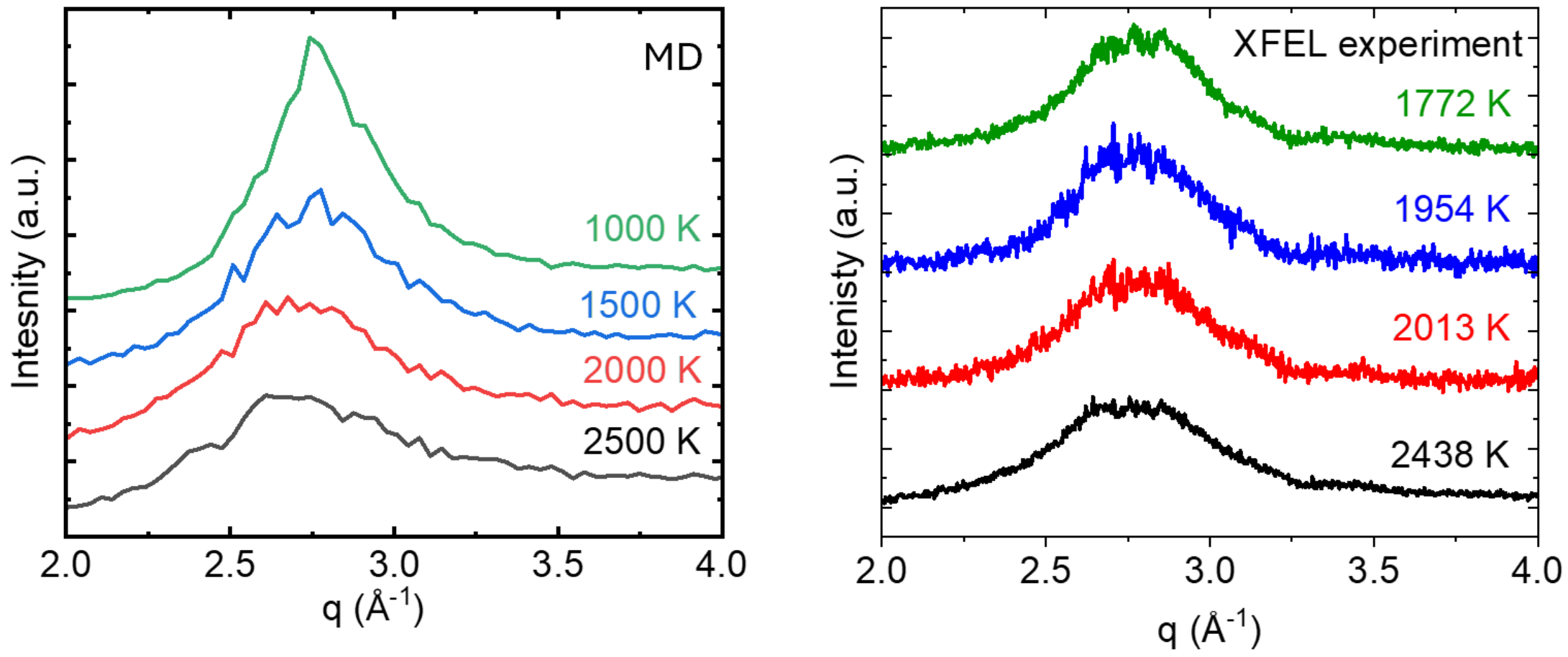

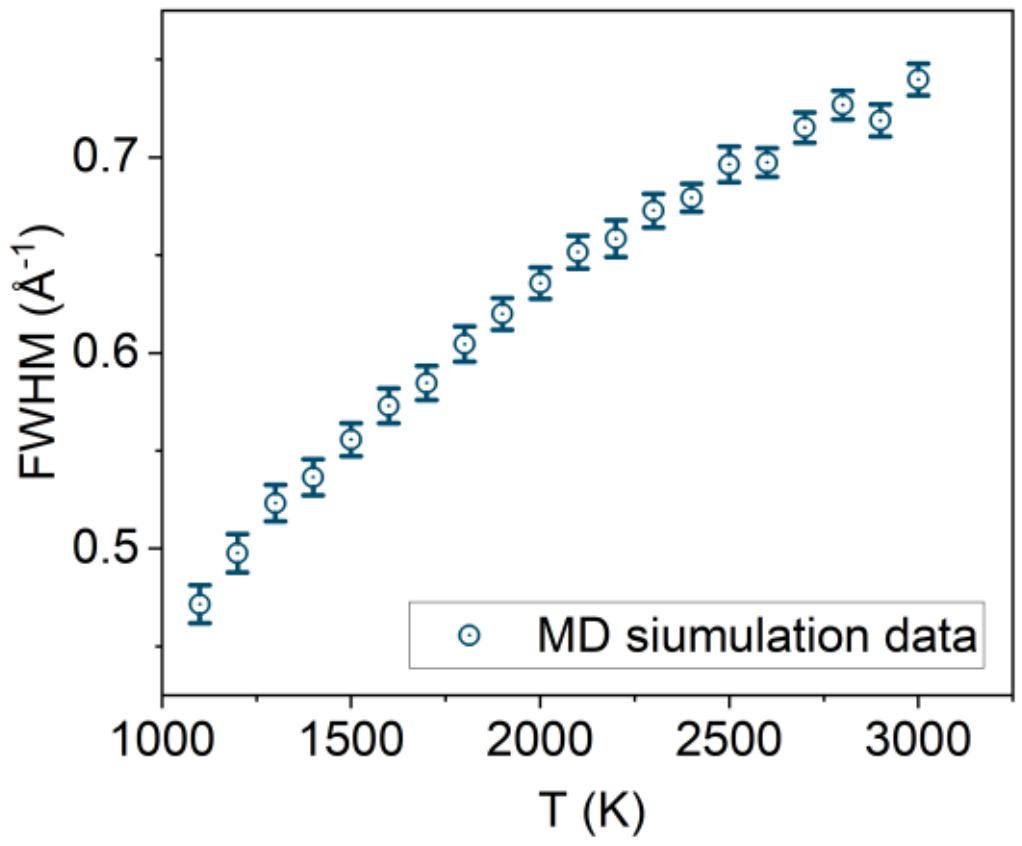

FWHM (Å$^{-1}$)
0.5
0.6
0.7
1000
1500
2000
2500
3000
T (K)
MD siumulation data